\documentclass [a4paper,12pt]{article}

\usepackage{anziamjedraft}

\usepackage {graphicx}
\usepackage{epsfig,amsmath,latexsym,amssymb}
\usepackage{bm}
\usepackage{booktabs}
\usepackage{multirow}

\title{Dependent default and recovery: \text{MCMC} study of downturn \text{LGD} credit risk model}

\author{Pavel V.~Shevchenko}
\address{CSIRO Mathematics, Informatics and Statistics, Locked Bag 17, North Ryde, 1670, Australia}
\mailto{pavel.shevchenko@csiro.au}

\author{Xiaolin Luo}
\address{CSIRO Mathematics, Informatics and Statistics, Locked Bag 17, North Ryde, 1670,
Australia}
\mailto{xiaolin.luo@csiro.au}

\date{\footnotesize{Working paper, 9 December 2011 }}

\begin{document}

\maketitle

\begin{abstract}
\noindent There is empirical evidence that recovery rates tend to go
down just when the number of defaults goes up in economic downturns.
This has to be taken into account in estimation of the capital
against credit risk required by Basel II to cover losses during the
adverse economic downturns; the so-called ``downturn LGD"
requirement. This paper presents estimation of the LGD credit risk
model with default and recovery dependent via the latent systematic
risk factor using Bayesian inference approach and Markov chain Monte
Carlo method. This approach allows joint estimation of all model
parameters and latent systematic factor, and all relevant
uncertainties. Results using Moody's annual default and recovery
rates for corporate bonds for the period 1982-2010 show that the
impact of parameter uncertainty on economic capital can be very
significant and should be assessed by practitioners.
\end{abstract}

\section{Introduction}
\label{sec:introductionords} Default and recovery rates are key
components of Loss Given Default (LGD) models used in some banks for
calculation of economical capital (EC) against credit risk. The
classic LGD model implicitly assumes that the default rates and
recovery rates are independent. Motivated by empirical evidence that
recovery rates tend to go down just when the number of defaults goes
up in economic downturns, Frye~\cite{Frye00},
 Pykhtin~\cite{Pykhtin} and
D\"{u}llmann and Trapp~\cite{DuellmannTr04} extended the classic
model to include dependence between default and recovery via common
systematic factor. These models have been suggested by some banks
for assessment of the Basel II ``downturn LGD" requirement
\cite{BaselPar468_2005}. The Basel II ``downturn LGD" reasoning is
that recovery rates may be lower during economic downturns when
default rates are high; and that a capital should be sufficient to
cover losses during these adverse circumstances. The extended models
represent an important enhancement of credit risk models used in
earlier practice, such as \emph{CreditMetrics}  and
\emph{CreditRisk+}, that do not account for dependence between
default and recovery.

Publicly available data provided by Moody's or Standard\&Poor's
rating agencies  are annual averages of defaults and recoveries.
These data are of limited size, covering a couple of decades at
most. As will be shown in this paper, due to the limited data size
the impact of the parameter uncertainty on capital estimate can be
very significant. None of the various studies, including the
extension works \cite{Frye00,Pykhtin,DuellmannTr04}, specifically
addressed the quantitative impact of parameter uncertainty.
Increasingly, quantification of parameter uncertainty and its impact
on EC has become a key component  of financial risk modeling and
management; for recent examples in operational risk and insurance,
see \cite{LuShDo07,PeShWu09}. This paper studies parameter
uncertainty and its impact on EC estimate in the LGD model, where
default and recovery are dependent via the latent systematic risk
factor. We demonstrate how the model can be estimated using Bayesian
approach and Markov chain Monte Carlo (MCMC) method. This approach
allows joint estimation of all model parameters and latent
systematic factor, and all relevant uncertainties.

\section{LGD Model}
Following \cite{DuellmannTr04,Frye00,Pykhtin}, consider a homogenous
portfolio of $J$ borrowers over a chosen time horizon. To avoid
cumbersome notation, we assume that the $j$th borrower has one loan
with principal amount $A_j$. The loss rate (loss amount relative to
the loan amount) of the portfolio due to defaults  is
\begin{equation}\label{LGD_model_eq}
L=\sum_{j=1}^J w_jL_j=\sum_{j=1}^J w_jI_j \max(1-R_j,0),
\end{equation}
\noindent where $w_j$ is the weight of loan $j$,
$w_j=A_j/\sum_{m=1}^JA_m$; $L_j$ is the loss rate of loan $j$ due to
potential default; $1-\max(1-R_j,0)$ is the recovery rate of loan
$j$ after default; $I_j$ is an indicator variable associated with
the default of loan $j$, $I_j=1$ if firm $j$ defaults, otherwise
$I_j=0$. In general $R_j$ is not the same as recovery rate since the
latter is subject to a cap of 1.

Denote the probability of default for firm $j$ by $p$, i.e.
$\mathrm{Pr}[I_j=1]=p$. Let  $C_j$ be an underlying latent  random
variable  (financial well-being) such that firm $j$ defaults if
$C_j<\Phi^{-1}(p)$, where $\Phi(\cdot)$  is the standard normal
distribution and $\Phi^{-1}(\cdot)$ is its inverse. That is, $I_j=1$
if $C_j<\Phi^{-1}(p)$ and $I_j=0$ otherwise.  The value $C_j$ for
each firm depends on a systematic risk factor $X$ and a firm
specific (idiosyncratic) risk factor $Z^C_j$ as

\begin{equation}\label{default_model_eq}
C_j=\sqrt{\rho}X+\sqrt{1-\rho}Z^C_j,
\end{equation}

\noindent where $Z^C_1,\ldots ,Z^C_J$ are all independent. Also, $X$
and $Z^C_j$ are assumed  independent and from the standard normal
distribution. Conditional on $X$, the financial conditions of any
two firms are independent. Unconditionally, $\rho$ is correlation
between financial conditions of two firms.

The studies \cite{DuellmannTr04,Frye00,Pykhtin} considered normal,
lognormal and logit-normal distributions for the recovery. It was
shown in \cite{DuellmannTr04} that EC estimates from these three
recovery models are very close to each other; the difference is
within $2\%$. In addition, statistical tests  favored the normal
distribution model. Thus we model the recovery rate as
\begin{equation}\label{recovery_model_eq}
 R_j=\mu +\sigma\sqrt{\omega}X+\sigma\sqrt{1-\omega}Z_j,
\quad \omega\in[0,1],
\end{equation}
\noindent where $X$ and $Z_j$ are assumed independent and from the
standard normal distribution. Also, $Z_j$ and $Z_j^C$ are assumed
independent too. The recovery and default processes are dependent
via systematic factor $X$.

\section{Economic Capital } It is common to define the EC for  credit risk
as a high quantile of the distribution of loss $L$, i.e.
\begin{equation}
 Q_q({\bm\theta })\equiv Q_q =  \inf \{z:\Pr [L > z|\bm\theta] \le 1 - q\}= \inf \{z:F_L(z|\bm\theta)\geq
 q\},
\label{eqn_quantile}
\end{equation}

\noindent where $q$ is a  quantile level; $F_L(z|\bm\theta)$ is
distribution function of the  loss $L$ with the density denoted as
$f_L(z|\bm\theta)$; and ${\bm\theta }=(p,\rho,\mu,\sigma,\omega)$
are model parameters.

The EC measured by the quantile $Q_q({\bm\theta })$ is a function of
 ${{\bm \theta }}$. Typically, given observations, the maximum likelihood estimators (MLEs)
 ${{\widehat{\bm\theta }}}$
are used as  point estimates for ${{\bm \theta }}$. Then, the  loss
density for the next time period is estimated as $f_L(z\vert
{{\widehat {\bm\theta }}})$ and its  quantile, $Q_q({{\widehat
{\bm\theta }}})$, is used for EC
 calculation. The distribution of $L$ is not
 tractable in closed form for an arbitrary portfolio. In this case
 Monte Carlo method can be
 used with the following logical steps.

~

 \noindent {\bf  Algorithm 1} ({\bf Quantile given
 parameters})

\noindent 1. Draw an independent sample from $\Phi(\cdot)$ for the
systematic factor $X$.

\noindent 2. For each $j$, draw $Z^C_j$ from $\Phi(\cdot)$;
calculate $C_j$ and $I_j$.

\noindent 3. For each $j$, draw $Z_j$ from $\Phi(\cdot)$ and find
$R_j=\mu +\sigma\sqrt{\omega}X+\sigma\sqrt{1-\omega}Z_j$.

\noindent 4. Find loss $L$ for the entire portfolio using (1), i.e.
a sample from $F_L(\cdot|\bm\theta)$.

\noindent 5. Repeat steps 1-4 to obtain $N$ samples of $L$.

\noindent 6. Estimate $Q_q({\bm \theta})$ using obtained samples of
$L$ in the standard way.

~

Bank loans are subject to the borrower specific risk and systematic
risk. In the case of a diversified portfolio with a large number of
borrowers, the idiosyncratic risk can be eliminated and the loss
depends on $X$ only.  Gordy~\cite{Gordy02} has shown that the
distribution of portfolio loss $L$ has a limiting form as
$J\rightarrow\infty$, provided that each weight $w_j$ goes to zero
faster than $1/\sqrt{J}$. The limiting loss rate $L^\infty$ is given
by the expected loss rate conditional on $X$
\begin{eqnarray} \label{eqn_infL0}
L^\infty&\equiv& L^\infty(X)=\sum_{j=1}^J w_jE[I_j|X]
E[\max(1-R_j,0)|X]=\Lambda(X) S(X),
\end{eqnarray}
\noindent where $\Lambda(X)=E[I_j|X]$ is the conditional
  probability of default of firm $j$ and
$S(X)=E[\max(1-R_j,0)|X]$ is the conditional expected value of loss
rate. That is, the distribution of $L^\infty$ is fully implied by
the distribution of $X$. Because $L^\infty(X)$ is a monotonic
decreasing function and $X$ is from the standard normal
distribution, the quantile of $L^\infty(X)$ at level $q$ can be
calculated as $Q^\infty_q =L^\infty\left(X=\Phi^{-1}(1-q)\right)$.
As  in \cite{DuellmannTr04}, we define EC of the diversified
portfolio loss distribution $L^\infty(X)$ as the $0.999$ quantile
\begin{eqnarray} \label{eqn_ec}
EC^{\infty}=Q^\infty_{0.999}
=L^\infty\left(\Phi^{-1}(0.001)\right)=\mathrm{PD}\times
\mathrm{LGD},
\end{eqnarray}
\noindent where $\mathrm{PD}= \Lambda(\Phi^{-1}(0.001))$  and
 $\mathrm{LGD}=S(\Phi^{-1}(0.001)$ are stressed probability of
default (\emph{stressed PD})
 and stressed loss given default (\emph{stressed LGD}) respectively.
Using (\ref{default_model_eq}), the conditional probability of
default is
\begin{equation}
\Lambda(X)=\Phi
\left(\frac{\Phi^{-1}(p)-\sqrt{\rho}X}{\sqrt{1-\rho}}\right ).
\label{eqn_lambda}
\end{equation}
\noindent Also, the expected conditional loss rate for the normally
distributed recovery rate model (\ref{recovery_model_eq}) is easily
calculated as
\begin{eqnarray}
S(X)&=&\int_{-\infty}^\infty
\max(1-\mu-\sigma\sqrt{\omega}X-\sigma\sqrt{1-\omega}z,0)f_N(z)dz \nonumber\\
&=&(1-\mu-\sigma\sqrt{\omega}X)\Phi(z_c)+\frac{\sigma\sqrt{1-\omega}}{\sqrt{2\pi}}e^{-z_c^2/2},
\end{eqnarray}
where $z_c={(1-\mu-\sigma\sqrt{\omega}X)}/{(\sigma\sqrt{1-\omega})}$
and $f_N(z)$ is the standard normal density. For the real data used
in this study, it can be well approximated as $S(X)\approx
 E[(1-R_j)|X]=1-\mu-\sigma\sqrt{\omega}X$.

\section{Likelihood}
Consider time periods $t=1,\ldots ,T$ (so that $T+1$ corresponds to
the next future year), where the following data of default and
recovery for a loan portfolio of $J_t$ firms are observed: $D_t$ --
the number of defaults in year $t$, and its realization is $d_t$;
$\Psi_t=D_t/J_t$ -- the  default rate year $t$, and its realization
is $\psi_t$; $\overline{R}_t=\sum_{j=1}^{D_t}R_j(t)/D_t$ -- the
average recovery rate in year $t$, where $R_1(t),\ldots ,R_{D_t}(t)$
are individual recoveries, and its realization is $\overline{r}_t$.
Also, the systematic factor $X$  corresponding to the time periods
is denoted as $X_1,\ldots,X_{T+1}$ and its realization is
$x_1,\ldots,x_{T+1}$. It is assumed that $X_1,\ldots,X_{T+1}$ are
independent and all idiosyncratic factors $(Z_j,Z_j^C)$
corresponding to the time periods are all independent.

\subsection{Exact Likelihood Function}
 The joint density of the number of defaults and average recovery
 rate ($D_t, \overline{R}_t)$ can be calculated by integrating out the
 latent variable $X_t$ for each $t$ as
\begin{equation}
f(d_t, \overline{r}_t)=\int  f(\overline{r}_t | d_t, x_t)f(d_t |
x_t)f_N(x_t)dx_t, \label{eqn_intX}
\end{equation}
where the conditional densities $f(d_t | x_t)$ and $f(\overline{r}_t
| d_t, x_t)$ are derived as follows.

Given $X_t=x_t$, all firms in a homogenous portfolio have the same
conditional default probability $\Pr[I_j(t)=1|X_t=x_t]=\Lambda(x_t)$
evaluated in (\ref{eqn_lambda}). Thus, the conditional distribution
of  $D_t=\sum_{j=1}^{J_t} I_j(t)$ is binomial
\begin{equation}
\label{eqn_bino}
 f(d_t |
x_t)=\Pr[D_t=d_t|X_t=x_t]=\binom{{J_t}}{{d_t}}
\left(\Lambda(x_t)\right)^{d_t}
\left(1-\Lambda(x_t)\right)^{J_t-d_t}.
\end{equation}
\noindent Often it can be well approximated by the normal
distribution $N(\mu_t, \sigma_t^2)$ with mean
$\mu_t=J_t\Lambda(x_t)$ and variance
$\sigma_t^2=J_t\Lambda(x_t)(1-\Lambda(x_t))$.

Conditional on $X_t=x_t$ and $D_t=d_t$; individual recoveries
$R_1(t),\ldots,R_{d_t}(t)$
 are independent and from
$N(\mu_r, \sigma_r^2)$ with $\mu_r=\mu +\sigma\sqrt{\omega}x_t$ and
$\sigma_r=\sigma\sqrt{1-\omega}$. Thus the average $\overline{R}_t$
is from $N(\mu_R, \sigma_R^2)$ with $\mu_R=\mu_r$ and
$\sigma_R^2=\sigma_r^2/d_t$, i.e.
\begin{equation}
\label{eqn_recovery_pdf}
 f(\overline{r}_t| d_t,
x_t)=\frac{1}{\sqrt{2\pi}\sigma_R} \exp \left
(-\frac{(\overline{r}_t-\mu_R)^2}{2\sigma_R^2}\right).
\end{equation}
\noindent If recovery distribution is different from normal, the
average  $\overline{R}_t$ can still be approximated by normal
distribution if $d_t$ is large (and variance is finite). Define the
data vectors $\bm{D}=(D_1,\ldots ,D_T)$ and
$\overline{\bm{R}}=(\overline{R}_1,\ldots ,\overline{R}_T)$, then
the joint likelihood function for data $\bm{D}$ and
$\overline{\bm{R}}$ is
\begin{equation}
\ell_{\bm{D},\overline{\bm{R}}}({\bm \theta})=\prod_{t=1}^{T}f(d_t,
\overline{r}_t). \label{eqn_joint_L}
\end{equation}
\noindent This joint likelihood function can be used to estimate
parameters $\bm{\theta}$ by MLEs maximizing this likelihood.
However, the likelihood involves numerical integration with respect
to the latent variables ${\bm X}$. It is difficult to accurately
compute these integrations, especially if the likelihood is used
within numerical maximization procedure. A straightforward and
problem-free alternative is to take Bayesian approach and treat
${\bm X}$ in the same way as other parameters, and formulate the
problem in terms of the likelihood conditional on ${\bm
\gamma}=({\bm \theta}, {\bm X})$. Then the required conditional
likelihood is easily calculated as
\begin{equation}
\ell_{\bm{D},\overline{\bm{R}}}({\bm \gamma})=\prod_{t=1}^T f(d_t |
x_t, {\bm \theta})f(\overline{r}_t | d_t, x_t, {\bm \theta}).
\label{eqn_joint_Lx}
\end{equation}
avoiding integration with respect to $\bm X$. Estimation based on
this likelihood will be discussed in detail in Section
\ref{sec:bayesian}.

\subsection{Approximate Likelihood and Closed-Form MLEs}
Assuming  a large number of firms in the portfolio, some
approximation can be justified to find MLEs for the likelihood
(\ref{eqn_joint_L}). We adopt approach from \cite{DuellmannTr04},
estimating the default process parameters $\bm{\theta}_D=(\rho, p)$
and systematic factor $\bm{X}$ first, and then fitting the recovery
parameters $\bm{\theta}_R=(\mu, \sigma, \omega)$.

Given $X_t$, the conditional default probability
$\Lambda_t=\Lambda(X_t)$ is a monotonic function of $X_t$; see
(\ref{eqn_lambda}). The density of $X_t$ is the standard normal,
thus the change of probability measure gives the density for
$\Lambda_t$ at $\Lambda_t=\lambda_t$:
\begin{equation}
f(\lambda_t |
\bm{\theta}_D)=\frac{1}{\sqrt{2\pi}}\exp\left(-\frac{x_t^2}{2}\right)\left|
\frac{dx_t}{d\lambda_t}\right|, \label{eqn_density_lambda}
\end{equation}
\noindent where $x_t$ is the function of $\lambda_t$, the inverse of
(\ref{eqn_lambda}),
\begin{equation}
x_t=\left({\Phi^{-1}(p)-\sqrt{1-\rho}\delta_t}\right)/{\sqrt{\rho}}.
\label{eqn_xt}
\end{equation}
\noindent Here, $\delta_t=\Phi^{-1}(\lambda_t)$. For year $t$ we
observe default rate $\Psi_t$ that for $J_t\rightarrow\infty$
approaches
 $\Lambda_t$. Therefore, the likelihood for
observed default rates $\bm{\psi}=(\psi_1,\ldots ,\psi_T)$ is
\begin{equation}
\ell_D(\bm{\theta}_D)=\prod_{t=1}^T{f(\lambda_t | \bm{\theta}_D)}
\label{eqn_likelihoodD}
\end{equation}
\noindent with $\delta_t=\Phi^{-1}(\psi_t)$. Maximizing
(\ref{eqn_likelihoodD})  gives  the following MLEs for $\rho$ and
$p$:
\begin{equation}
\hat{\rho}=\frac{\sigma_{\bm{\delta}}^2}{1+\sigma_{\bm{\delta}}^2},
\quad
\hat{p}=\Phi\left(\frac{\overline{\delta}}{\sqrt{1+\sigma_{\bm{\delta}}^2}}\right),
\label{eqn_rho_p}
\end{equation}
\noindent where $\overline{\delta}=\sum_{t=1}^{T}\delta_t/T$
 and $\sigma_{\bm{\delta}}^2=\sum_{t=1}^{T}(\delta_t-\overline{\delta})^2/T$.
 The
factor $X_t$ is then estimated using (\ref{eqn_xt}) with default
parameters $(p, \rho)$ replaced by MLEs as
\begin{equation}
\hat{x}_t=\left({\Phi^{-1}(\hat{p})-\sqrt{1-\hat{\rho}}\delta_t}\right)/{\sqrt{\hat{\rho}}}.
\label{eqn_xt2}
\end{equation}
Given $X_t$ and $D_t$, the average recovery rate $\overline{R}_t$
 is from  $N(\mu_R,\sigma^2_R)$ with mean
$\mu_R=\mu+\sigma\sqrt{\omega}X_t$ and variance
$\sigma^2_R=\sigma^2(1-\omega)/d_t$. Thus the likelihood for $T$
observations of the average recovery rate
$\overline{\bm{r}}=(\overline{r}_1,\ldots ,\overline{r}_T)$ is
\begin{equation} \label{eqn_likelihoodR}
\ell_{\overline{\bm{R}}}(\bm{\theta}_R, \bm{x})=\prod_{t=1}^T
\sqrt{\frac{d_t}{2\pi\sigma^2(1-\omega)}}\exp\left(
-\frac{d_t(\overline{r}_t-\mu-\sigma\sqrt{\omega}x_t)^2}{2\sigma^2(1-\omega)}\right).
\end{equation}
\noindent D\"{u}llmann and Trapp~\cite{DuellmannTr04} estimate
$\bm{\theta}_R$ by MLEs via maximization of (\ref{eqn_likelihoodR})
with respect to $\bm{\theta}_R$, where $x_t$ is replaced with
$\hat{x}_t$. Due to numerical difficulties with maximization, they
estimate $\sigma$ by the historical volatility $\widehat{\sigma}_h$
of the recovery rate $\overline{r}_t$. However, re-parameterizing
with $\sigma_1=\sigma\sqrt{\omega}$ and
$\sigma_2=\sigma\sqrt{1-\omega}$, we derive the following
closed-form solutions for MLEs of $(\mu, \sigma, \omega)$:
\begin{equation} \label{eqn_sigma1}
\widehat{\sigma}_1=\frac{\left(\sum_t d_t\overline{r}_tX_t
\right)\left(\sum_t d_t\right) - (\sum_t d_t\overline{r}_t )(\sum_t
d_tX_t ) }{\left(\sum_t d_tX_t^2\right)\left(\sum_t d_t\right)
-\left(\sum_t d_tX_t \right)^2 },
\end{equation}
\begin{equation} \label{eqn_mu}
\widehat{\mu}=\frac{(\sum_t d_t\overline{r}_tX_t ) - (\sum_t d_t
X_t^2 )\widehat{\sigma}_1 }{\sum_t
d_tX_t},\quad\widehat{\sigma}_2=\sqrt{\frac{1}{T}\sum_t d_t
(r_t-\widehat{\mu}-\widehat{\sigma}_1 X_t)^2 },
\end{equation}
\begin{equation} \label{eqn_omega}
\widehat{\omega}=\frac{\widehat{\sigma}_1^2}{\widehat{\sigma}_1^2+\widehat{\sigma}_2^2},\quad
\widehat{\sigma}=\sqrt{\widehat{\sigma}_1^2+\widehat{\sigma}_2^2 }.
\end{equation}

\section{Bayesian Inference and MCMC}
\label{sec:bayesian} The parameters ${{\bm \theta }}$ are unknown
and it is important to account for this uncertainty when the capital
is estimated. A standard frequentist approach to estimate this
uncertainty is based on limiting results of normally distributed
MLEs for large datasets. We take Bayesian approach, because dataset
is small and  parameter uncertainty distribution is very different
form normal. From a Bayesian perspective, both parameters ${\bm
\theta }$ and latent factor ${\bm X}$ are random variables. Given a
\textit{prior} density $\pi ({\bm \gamma })$ and a data likelihood
$\pi ({\rm {\bf y}}\vert {\bm \gamma })=\ell_{\bm{Y}}(\bm{\gamma})$,
where ${\bm \gamma}=({\bm \theta }, {\bm X})$ and $\bm{Y}$ is data
vector, the density of ${\bm \gamma }$ conditional on ${{\bm
Y}=\bf{y}}$ (\textit{posterior} density) is determined by the Bayes
theorem
\begin{equation}
\pi ({\bm \gamma }\vert {\rm {\bf y}}) \propto {\pi ({\rm {\bf
y}}\vert {\bm \gamma })\pi ({\bm \gamma })}. \label{eqn_bayes}
\end{equation}
The posterior can then be used for predictive inference and analysis
of the uncertainties. There are many useful texts on Bayesian
inference; e.g. see \cite{Rob01}; for recent examples in operational
risk and insurance, see \cite{Shevchenko2011,PeShWu09,PeShWu09b}.

The explicit evaluation of posterior (\ref{eqn_bayes}) is often
difficult and one can use MCMC method to sample from the posterior.
In particular,  MCMC allows to get samples of ${\bm \theta }$ and
${\bm X}$ from the joint posterior $\pi ({\bm \theta }, {\bm X}\vert
{\rm {\bf y}})$. Then taking samples of ${\bm \theta }$ marginally,
we can get the posterior for model parameters $\pi ({\bm \theta
}\vert {\rm {\bf y}})$, i.e. effectively integrating out the latent
factor $\bm X$. Similarly, taking samples of ${\bm X}$ marginally,
we get the posterior for systematic factor  $\pi ({X_t }\vert {\rm
{\bf y}})$. Posterior mean is commonly used point estimate. We adopt
component-wise Metropolis-Hastings algorithm for sampling from
posterior $ \pi ({\bm \gamma}\vert {\rm {\bf y}})$, following the
same procedure as in \cite{ShTem09,PeShWu09}. Other MCMC methods
such as the univariate slice sampler utilized in \cite{PeShWu09b}
can also be used. For numerical efficiency, we work with
 parameter $\Phi^{-1}(p)$ .
Also, we assume a uniform prior for all parameters  and the standard
normal distribution as the prior for $X_1,\ldots ,X_T$. The only
subjective judgement we bring to the prior is the lower and upper
bounds of the parameter values
$$\Phi^{-1}(p)\in (-10,10),\quad\rho\in (0,1),\quad \mu\in(0,1),\quad\sigma\in
(0.01,  1.0),\quad\omega\in (0, 1).$$ The parameter support range
should be sufficiently large so that the posterior is implied mainly
by the observed data.
 We checked that
 an increase in parameter bounds did not lead to material difference in results.

The starting value of the chain for the $k$th component is set to a
uniform random number drawn independently from the support
$(a_k,b_k)$. In the single-component Metropolis-Hastings algorithm,
we adopt a Gaussian density (truncated below $a_k$ and above $b_k $)
for the proposal density. For each component the variance parameter
of proposal was pre-tuned and adjusted so that the acceptance rate
is close to 0.234 (optimal acceptance rate for $d$-dimensional
target distributions with iid components as shown in \cite{RoRo01}).
The chain is run for $100,000$ posterior samples (after $20,000$
``burn-in'' samples).

\section{Bayesian Capital Estimates} As
discussed in \cite{Shevchenko08a}, Bayesian methods are particularly
convenient to quantify parameter uncertainty and its impact on
capital estimate. Under the Bayesian approach,  the full predictive
density (accounting for parameter uncertainty) of the next time
period loss $L_{T + 1}$, given  data ${\rm\bf{Y}}=\bf y$, is
\begin{equation}
\label{FullPredPDF_eq} f_{L_{T+1}}(z\vert {\rm {\bf y}}) = \int
{f_{L_{T+1}}(z\vert {{\bm\theta }}) \pi ({{\bm\theta }}\vert {\rm
{\bf y}})d{{\bm\theta }}},
\end{equation}
\noindent assuming that, given ${{\bm \Theta }}$, $L_{T + 1}$ and
${\rm {\bf Y}}$ are independent. Its quantile,
\begin{equation}
\label{QuantileFullPred_eq} Q_q^P =  \inf \{z:\Pr [L_{T + 1}
> z\vert {\rm {\bf Y}}] \le 1 - q\},
\end{equation}
\noindent can be used as a risk measure for EC. The procedure for
 simulating $L_{T+1}$ from (\ref{FullPredPDF_eq}) and calculating $Q_q^P$
is simple: 1) Draw a sample of ${\bm\theta }$ from the posterior
  $\pi({\bm\theta } | {\bm y})$, e.g. using MCMC;
2) Given  ${\bm\theta }$, simulate loss $L$ following  steps 1-4 in
Algorithm 1; 3) Repeat steps 1-2 to obtain $N$ samples of $L$; 4)
Estimate $Q^P_q$ using samples of $L$ in the standard way.

Another approach under a Bayesian framework to account for parameter
uncertainty is to consider a quantile $Q_q ({{\bm \Theta }})$ of the
 loss density $f(\cdot\vert {{\bm \Theta }})$,
\begin{equation}
\label{DistrOfQuantile_eq} Q_q ({{\bm \Theta }}) =  \inf \{z:\Pr
[L_{T + 1} > z\vert {{\bm \Theta }}] \le 1 - q\}.
\end{equation}
\noindent Given that ${{\bm \Theta }}$ is distributed as $\pi ({{\bm
\theta }}\vert {\rm {\bf y}})$, one can find the associated
distribution of $Q_{q} ({{\bm \Theta }})$, form a predictive
interval to contain the true quantile value with some probability
and argue that the conservative estimate of the capital  accounting
for parameter uncertainty should be based on the upper bound of the
 interval. However it might be difficult to
justify the choice of the confidence level for the interval. The
procedure to obtain the posterior distribution of quantile $Q_{q}
({{\bm \Theta }})$ is simple: 1) Draw a sample of ${\bm\theta }$
from the posterior $\pi({\bm\theta } | {\bf y})$, e.g. using MCMC;
2) Compute $Q_q=Q_q({\bm \theta})$ using e.g. Algorithm 1; 3) Repeat
steps 1-2 to obtain $N$ samples of $Q_q ({{\bm \Theta }})$. For
limiting case of large number of borrowers, Step 2 can be
approximated by a closed-form formula.

The extra loading for EC due to parameter uncertainty can be
formally defined as the difference between the quantile of the full
predictive distribution accounting for parameter uncertainty
$Q_{0.999}^P$ and posterior mean of $Q_{0.999} ({{\bm \Theta }})$,
i.e. $Q_{0.999}^P-E[Q_{0.999} ({{\bm \Theta }})]$.

\section{Results using Moody's data}
\label{sec:mylabel1} 
\begin{figure}
\centerline{\includegraphics[scale=0.7]{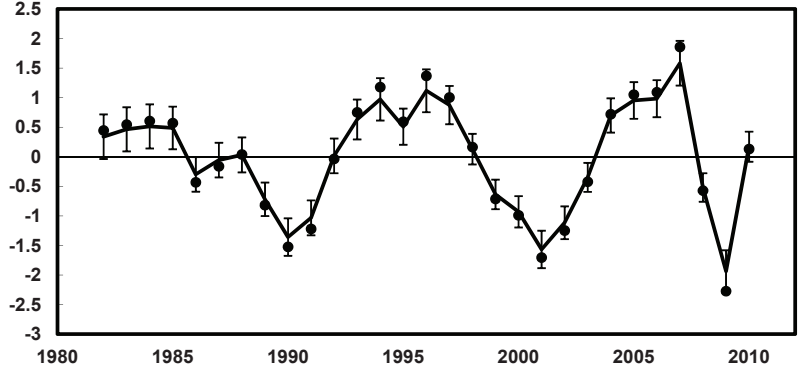}} \caption{MLE
(dots) and posterior mean (solid line) of systematic factor $X_t$.
Error bars correspond to posterior standard deviation of $X_t$.}
\label{fig6}
\end{figure}

Using historical data for the overall corporate
default and recovery rates over 1982-2010 from Moody's report
\cite{Moody2011}, we fit the model using MCMC and MLEs. Table
\ref{parameter_estimates_tab} shows posterior summary and MLE for
the model parameters (the coefficient of variation, CV, is defined
as
 the ratio of standard deviation to the mean). Significant kurtosis and
 positive skewness in most parameters
indicate that Gaussian approximation for parameter uncertainties is
not appropriate. Also, all MLEs are within one standard deviation
from the posterior mean. The posterior mean of systematic factor
$X_t$ for 2009 is about -2.27, which corresponds to approximately
$99\%$ quantile level  of the diversified portfolio. This maximum
negative systematic factor for 2009 is the consequence of the
disastrous 2008 when the bankruptcy of Lehman Brothers occurred.
Comparison of MLE and posterior mean for latent factor $\bm X$
  is shown in Figure
\ref{fig6}.

\begin{table}[htbp]
\begin{center}
\caption{MLE and MCMC posterior statistics of the model parameters.}
\begin{tabular*}
{0.95\textwidth}{cccccccc}
\toprule item & MLE  & Mode &  Mean &  Stdev & Skewness & Kurtosis & CV \\
 \midrule
 $p$ & 0.0167  &  0.0177  & 0.0179 & 0.0028 & 0.812 & 4.62 & 0.154  \\

 $\rho$ &  0.0635 & 0.141 & 0.0815  & 0.024  & 1.01 & 4.35 & 0.286  \\

  $\mu$ & 0.411  & 0.439 & 0.414  & 0.022 & 0.309 & 3.19 & 0.055  \\
$\omega$ & 0.0192 & 0.0717 & 0.031   & 0.016 &  1.24 & 5.39 & 0.51  \\
$\sigma$ & 0.499 & 0.449 & 0.502   & 0.070 &  0.588 & 3.63 & 0.140  \\
 \bottomrule
\end{tabular*}
\label{parameter_estimates_tab}
\end{center}
\end{table}

The MCMC predictions on stressed PD, LGD and EC in comparison with
corresponding MLEs
 are shown in  Table
\ref{PD_LGD_EC_estimates_tab}. MLE for EC is $35\%$ lower than the
posterior mean, $24\%$ lower than the posterior median and more than
$50\%$ lower than the 0.75 quantile of the posterior for EC. The
uncertainty in the posterior of EC is large, CV is about 34.5\%;
also note a large difference between the 0.75 and the 0.25 quantiles
of EC posterior.  Underestimation of EC by MLE in comparison with
posterior estimates is significant due to large parameter
uncertainty and large skeweness in EC posterior. Also, we get the
following results for the 0.999 quantile ${Q}^P_{0.999}$ of the full
predictive loss density for portfolios with different number of
borrowers $J$: $Q^P_{0.999}=(0.1454, 0.1092, 0.1026, 0.1026)$ for
$J=(50,  500, 5000,  \infty)$ respectively. The
 diversification effect when $J$ increases is
 evident. In particular, $Q^P_{0.999}$ at $J=500$
 is about $25\%$ lower than the case at $J=50$; and
for $J=5000$ is virtually the same as for the limiting case
$J=\infty$. Note that ${Q}^P_{0.999}$ at $J=\infty$ is about 50\%
larger than MLE for EC; and about 15\% larger than the posterior
mean of $Q^{\infty}_{0.999} ({\bm \Theta })$. The 15\% impact of
parameter uncertainty on EC gives indication that $1982-2010$
dataset is long enough for a more or less confident use of the model
for capital quantification. Of course, a formal model validation
should be performed before final conclusion.
\begin{table}[htbp]
\begin{center}
\caption{MLE and MCMC posterior statistics for PD, LGD and EC.}
\begin{tabular*}
{0.9\textwidth}{cccccccc}
\toprule item &  MLE  &  Mean & Stdev &   0.25Q & 0.5Q & 0.75Q &  CV \\
 \midrule
 PD &   0.0819   & 0.103 & 0.029 &  0.0825 & 0.0968 & 0.116  & 0.288  \\
  LGD &  0.803    & 0.847 & 0.0562 &   0.808 & 0.841 & 0.880 & 0.066  \\
 EC &  0.0657      & 0.0888 & 0.031  & 0.0672 & 0.0814 & 0.101 & 0.345  \\
 \bottomrule
\end{tabular*}
\label{PD_LGD_EC_estimates_tab}
\end{center}
\end{table}


\section{Conclusion}
Presented methodology allows joint estimation of the model
parameters and latent systematic risk factor in the well known LGD
model via Bayesian approach and MCMC method. This approach allows an
easy calculation of the full predictive loss density
$f_{L_{T+1}}(\cdot|\bf y)$ accounting for parameter uncertainty;
then the economic capital can be based on the high quantile of this
distribution. Given small datasets typically used to fit the model,
the parameter uncertainty is large and the posterior is very
different from the normal distribution indicating that Gaussian
approximation for parameter uncertainties (typically used under the
frequentist maximum likelihood approach assuming large sample limit)
is not appropriate.

Due to data limitation, we assumed homogeneous portfolio and thus
the results should be treated as illustration. However, the results
demonstrate that the extra capital to cover parameter uncertainty
can be  significant and should not be disregarded by practitioners
developing LGD models. The approach can be extended to deal with
non-homogeneous portfolios, more than one latent factor and mean
reversion in the systematic factor. It should not be difficult to
incorporate macroeconomic factors as in \cite{Scheule05}.

{\footnotesize{
\bibliography{bibliography}
\bibliographystyle{plain}
}}
\end{document}